\pdfoutput=1
\documentclass{sig-alternate-2013}
\usepackage[
  pass,
]{geometry}
\usepackage{url}
\permission{\copyright 2017 International World Wide Web Conference Committee \\ (IW3C2), published under Creative Commons CC BY 4.0 License.}
\conferenceinfo{WWW'17 Companion,}{April 3--7, 2017, Perth, Australia.}
\copyrightetc{ACM \the\acmcopyr}
\crdata{978-1-4503-4914-7/17/04. \\
http://dx.doi.org/10.1145/3041021.3053366 \\
\includegraphics{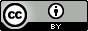}}

\begin{document}

\title{Building Automated Vandalism Detection Tools for Wikidata}
\numberofauthors{3}

\author{
\alignauthor Amir Sarabadani\\
\affaddr{Wikimedia Deutschland}\\
\affaddr{Tempelhofer Ufer 23/24}\\
\affaddr{10963 Berlin, Germany}\\
\email{Ladsgroup@gmail.com}
\alignauthor Aaron Halfaker\\
\affaddr{Wikimedia Research}\\
\affaddr{149 New Montgomery Street}\\
\affaddr{San Francisco, USA}\\
\email{ahalfaker@wikimedia.org}
\alignauthor Dario Taraborelli\\
\affaddr{Wikimedia Research}\\
\affaddr{149 New Montgomery Street}\\
\affaddr{San Francisco, USA}\\
\email{dtaraborelli@wikimedia.org}
}

\maketitle
\begin{abstract}
Wikidata, like Wikipedia, is a knowledge base that anyone can edit.  This open collaboration model is powerful in that it reduces barriers to participation and allows a large number of people to contribute. However, it exposes the knowledge base to the risk of vandalism and low-quality contributions.  In this work, we build on past work detecting vandalism in Wikipedia to detect vandalism in Wikidata.  This work is novel in that identifying damaging changes in a structured knowledge-base requires substantially different feature engineering work than in a text-based wiki like Wikipedia.  We also discuss the utility of these classifiers for reducing the overall workload of vandalism patrollers in Wikidata.  We describe a machine classification strategy that is able to catch 89\% of vandalism while reducing patrollers' workload by 98\%, by drawing lightly from contextual features of an edit and heavily from the characteristics of the user making the edit.
\end{abstract}

\section{Introduction}
Wikidata (www.wikidata.org) is a free knowledge base that everyone can edit. It is a collaborative project aiming to produce a high quality, language-independent, open-licensed, structured knowledge base.  Like Wikipedia, the project is open to anyone willing to contribute productively.  This also opens Wikidata to potentially damaging/disruptive contributions. In order to combat such intentional damage, volunteer patrollers work to review changes to the database after they are saved. At a rate of about 80,000 human edits and 200,000 automated edits per day (as of February 2016), though, the task of reviewing every single edit would be daunting even for a very large pool of patrollers. Recently, substantial concerns have been raised about the quality and accuracy of Wikidata's statements \cite{kolbe:whither}, and therefore, the long-term viability of the project. These concerns call for the design of scalable quality control processes.

Similar concerns about quality control have been raised about Wikipedia in the past \cite{giles:internet}.  Studies of Wikipedia's quality have shown that, even at large scale and with open permissions, a high-quality information resource can be maintained \cite{giles:internet, stvilia:information}.  One of the key technologies that let Wikipedia maintain quality efficiently at scale is the use of machine classifiers for detecting vandalism edits. These technologies allow the massive feed of daily changes to be filtered down to a small percentage that is most likely to actually be vandalism, substantially reducing the workload of patrollers \cite{geiger:levee, geiger:work}.  These semi-automated support systems also substantially reduce the amount of time that an article in Wikipedia remains in a vandalized state \cite{geiger:levee}.  The study of vandalism detection in Wikipedia has seen substantial development as a field in the scholarly literature, to great benefit of the project \cite{wang:got, harpalani:language, adler:detecting, adler:wikipedia}.

In this study, we extend and adapt methods from the Wikipedia vandalism detection literature to Wikidata's structured knowledge base.  In order to do so, we develop novel techniques for extracting signal from the types of changes that editors make to Wikidata's \textit{items}. But unlike this past literature, we focus our evaluation on the key concerns of Wikidata patrollers who are tasked with reviewing incoming edits for vandalism: reducing their workload. We show that our machine classifier can be used to reduce the amount of edits that need review by up to 98\% while still maintaining a recall of 89\% using an off-the-shelf implementation of a Random Forest classifier \cite{breiman:random}\footnote{http://scikit-learn.org/}.

\subsection{Wikidata in a nutshell}
Wikidata consists of mainly two types of entities: \textit{items} and \textit{properties}. \textit{Items} represent define-able \textit{things}. Since Wikidata is intended to operate in a language-independent way, each \textit{item} is uniquely identified by a number prefixed with the letter ``Q''.  \textit{Properties} describe a data value of a statement that can be predicated of an item.  Like items, properties are uniquely identified by a number prefixed with the letter ``P''.

Each \textit{item} in Wikidata consists of five sections.

\begin{description}
\item[Labels] a name for the item (unique per language)
\item[Descriptions] a short description of the item (unique per language)
\item[Aliases] alternative names that could be used as a label for the item (multiple aliases can be specified per language)
\item[Statements] \textit{property} and data value pairs such as country of citizenship, gender, nationality, image, etc.  Statements can also include \textit{qualifiers} (which include sub-statements like the date of a census for a population count) and sources (like Wikipedia, Wikidata demands reliable sources for its data)
\item[Site links] links to Wikipedia and other Wikimedia projects (such as Wikisource\footnote{https://wikisource.org}) that reference the item.
\end{description}

For example, the item representing the city of San Francisco (Q62) contains the following statement: (P190, Q90).  P190 is a property described in English as ``sister city'' and Q90 is an item for the city of ``Paris, France''.  This statement represents the fact that San Francisco (Q62) has a sister city (P190) named Paris (Q90). Using so-called SPO triplets, standing for subject-predicate-object, as a mean to store knowledge is a common practice in knowledge bases such as Freebase.

\section{Related work}
The subject of quality in open production has been extensively studied in the open text editing contexts like Wikipedia, but comparatively little study has been done in open structured data editing contexts. In this section, we'll provide an overview of some of the most relevant work exploring quality in open contexts like Wikipedia and Wikidata.

Wikidata and Wikipedia operate in a common context: they are supported by the Wikimedia Foundation\footnote{https://wikimediafoundation.org}; virtually all of Wikidata users are also active in Wikipedia and/or other Wikimedia wikis; Wikidata, like Wikipedia, is powered by MediaWiki software, but Wikidata uses the ``wikibase'' extension\footnote{http://wikiba.se/} to manage structured data. Thus, damage detection in Wikipedia is closely related to vandalism detection in Wikidata projects. However, there are also open contribution structured data projects where quality and vandalism detection have been a focus of scholarly inquiry.

\subsection{Quality in Wikipedia}
Quality in Wikipedia has been studied so extensively that we can't give a fair overview of all related work, so here, we provide a limited overview of the work that is related to quality prediction and editing dynamics.

Stvilia et al. built the first automated quality prediction models for Wikipedia that was able to distinguish between Featured (highest quality classification) and non-Featured articles \cite{stvilia:assessing}.  Warncke-Wang et al. extended this work by showing that the features used in prediction could be limited to \textit{actionable} characteristics of articles in Wikipedia and maintain a high level of fitness \cite{warncke:tell} and used these predictions in task routing.

Kittur et al. explored the process by which articles improve most efficiently and found that articles with a small group of highly active editors and a large group of less active editors were more likely to increase in quality than articles whose editors contributed more evenly \cite{kittur:harnessing}. They argued that this is due to the lower coordination cost when few people are primarily engaged in the construction of an article.  Arazy et al. challenged the conclusions of Kittur et al. by showing a strong correlation between diversity of experience (global inequality) between editors who are active and positive changes in article quality \cite{arazy:determinants}. The visibility of articles in Wikipedia seems to be critical to their development.  Schneider et al., showed that hiding newly created articles from Wikipedia readers in a drafting space substantially reduced the overall productivity of editors in Wikipeida \cite{schneider:accept}.

Detecting vandalism in Wikipedia using machine learning classifiers has been an active area since 2008 \cite{smets:automatic}. There are generally two types of damage detection problems discussed in the literature: \textit{realtime} and \textit{post-hoc}.  The \textit{realtime} framing of damage detection imagines the classifier supporting patrollers by helping them find vandalism shortly after it has happened.  The \textit{post-hoc} framing of damage detection imagines the classifier being used long after an edit has been saved (and potentially reverted by patrollers).  Since the post-hoc framing allows the model to take advantage of what happens to a contribution after it is saved (e.g. that it was reverted), these classifiers are able to attain a much higher level of fitness than \textit{realtime} classifiers that must make judgement before a human has responded to an edit.  However, the utility of a post-hoc classifier is at best hypothetical while \textit{realtime} classifiers have become a critical infrastructure for Wikipedia patrollers \cite{geiger:work, geiger:levee}.  Geiger et al. discussed the ``distributed cognition'' system that formed through the integration of counter-vandalism tools that use machine classification and social practices around quality control \cite{geiger:work}.  Geiger et al. showed in a follow-up work that, when systems that use automated vandalism detection go offline, vandalism is not reverted as quickly which results in twice as many views of vandalized articles \cite{geiger:levee}.

Substantial effort has been put into developing high signal features for vandalism classifiers.  Adler et. al were able to show substantial gains in model fitness when including user-reputation metrics as features \cite{adler:detecting}. Among several other metrics, they primarily evaluated the fitness of their classifier using the area under the receiver operating characteristic curve (ROC-AUC) and were able to attain 93.4\%. Other researchers explored the use of stylometric features and were able to attain an ROC-AUC of 92.9\% \cite{wang:got,harpalani:language}.  West et al. explored spacio-temporal features and built and evaluated models predicting vandalism over only anonymous (not logged-in) user edits because those are the editors from which most vandalism (``offending edits'' in West et al.'s terminology) originate \cite{west:detecting}.  Adler et al. continues this work by comparing all of these feature extraction strategies/models and combines them to attain an ROC-AUC of 96.3\% \cite{adler:wikipedia}.  They continue to call for a focus on the area under the precision-recall curve (PR-AUC) instead of the ROC-AUC since it affords more discriminatory power between the overall fitness of models in the context of a low prevalence prediction problem (few positive examples -- as is the case with vandalism in Wikipedia).

\subsection{Quality in Wikidata}
Like Wikipedia, Wikidata is based on the MediaWiki software which provides several means for tracking and reviewing changes to content. For example, watchlists\footnote{https://en.wikipedia.org/wiki/Special:Watchlist} allow editors to be notified about changes made to items and properties that they are interested in. The recentchanges feed\footnote{https://en.wikipedia.org/wiki/Special:RecentChanges} provides an interface for reviewing all changes that have been made to the knowledge base. Wikidata also uses tools related to its own quality demands. Most notably, ``Constraint violation reports''\footnote{https://www.wikidata.org/wiki/WD:CV} is a dynamic list of possible errors in statements that is generated using predefined rules for properties.  For example, if a feline is stated to be a spouse of a human being, that's likely to need review. Other tools such as Kian\footnote{https://github.com/Ladsgroup/Kian} also expose possible errors in Wikidata by comparing data in Wikidata with extracted values from Wikipedia. Despite all of the efforts on quality control in Wikidata, still concerns have been raised regarding Wikidata reliability. For example Kolbe \cite{kolbe:whither} calls into question whether volunteers will ever be able to verify and source the statements in Wikidata.

Regarding vandalism detection Heindorf et al. have studied on the demography of vandalism in Wikidata \cite{heindorf:towards} showing interesting dynamics in how and who vandalizes Wikidata. For example, most of the vandals in Wikidata had previously vandalized Wikipedia. \cite{heindorf:towards} As far as we can tell, our work is the first published about a vandalism detection classifier for Wikidata.

\subsection{Quality in other structured data repositories}
There have been several research projects conducted on damage detection in knowledge bases. Most notably, Tan et al. \cite{tan:trust} worked on detecting correctness of data added to freebase\footnote{Freebase, Google's knowledge base, is shut down in favor of Wikidata.}. They assumed that, if a statement can survive for four weeks, it's probably a good contribution.  They also showed the ratio of correct statements added by a user is not predictive in determining the correctness of future statements, but by defining the area of expertise for each user, it's possible to make proper predictions. This work doesn't apply directly to our work exploring vandalism detection in Wikidata because they formalize the problem in terms of correctness of data while our aim is detecting vandalism. Nies et al. \cite{neis:towards} have done a research regarding vandalism in OpenStreetMap (OSM). OSM, like Wikidata, is an open structured database but unlike our work, they did not draw from the substantial history of vandalism detection in Wikipedia. Also, they did not use machine learning. Their method is poorly described ``rule-based'' scoring system and would be difficult to reproduce. In our work, we draw extensively from past work building high fitness vandalism detection models for Wikipedia. We use training and testing strategies that are intended to be straightforward to replicate.  We've adopted standard metrics from the Wikipedia vandalism detection literature and supplement our own intuitive evaluation metric (filter-rate) that correspond to real effort saved for Wikidata patrollers.

\section{Methods}
Designing damage detection classifiers that can be effectively used by Wikidata patrollers requires two conditions. First, the classifier should be able to respond and classify edits in a timely manner (i.e. within a few seconds); reviewing large sets of edits would be unfeasible with longer response times. On average, two edits are made by human editors in Wikidata every second. Given this high edit rate, using post-hoc features (such as the time an edit stays without being reverted) is undesirable in a production environment. Second, two distinct use cases need to be supported: auto-reverting of edits by bots and triaging edits to be reviewed by humans. In the first use case, the classifier is expected to have a high level of confidence, for instance a 90-99\% precision. In this case, the classifier will catch obvious vandalisms (e.g. blanking of a statement) but a higher recall would be helpful. In the second use case (human review), the classifier is expected to have a high recall: low precision can be tolerated, however it should not be too low so that in practice it classifies all edits as potential vandalism.

In order to build a classifier usable by Wikidata users, we leverage the Wikimedia Labs infrastructure hosted by the Wikimedia Foundation\footnote{https://www.mediawiki.org/wiki/Wikimedia\_Labs}. We rely on a service called ORES (for Objective Revision Evaluation Service), which can host machine learning classifiers for all projects by the Wikimedia Foundation, including Wikipedia and Wikidata. ORES accepts two methods of scoring edits: a single edit mode and a batch mode. We tested ORES response time by testing 1,000 randomly sampled edits. Response time in single edit mode varies between 0.0076 and 14.6 seconds with a mean of 0.66 seconds and median of 0.53 seconds. In batch mode with sets of 50 edits, response time falls between 0.56 and 13.9 seconds with a mean of 6.23 seconds and median of 5.58 seconds. Analyzing the response time in single edit mode, two peaks are noticeable: the first peak is around 0.3 seconds and a second peak is around 0.55 seconds.

\subsection{Building a corpus}
While there has been substantial work done in the past to build a high quality vandalism corpus for Wikipedia \cite{potthast:crowdsourcing}, no such work has been done for Wikidata.  Research by Heindorf et al. \cite{heindorf:towards} was intended to build such a corpus, but their method (matching edit comments for the use of specific tool) is inapplicable as it mislabels a substantial amount of edits.  They also assume that patrollers only use one of two methods available in the editing interface to revert vandalism: ``rollback'' and ``restore''.  Their qualitative analysis shows that 86\% of rollbacked edits and 62\% of edits reverted using the restore feature were vandalism.  If a classifier trained using this limited corpus proved useful for predicting all cases of vandalism regardless of the reverting method, then it would perform poorly when trying to predict true vandalism edits that were mislabeled in the corpus (i.e., reverted by ways other than the ``rollback'' method).  In a second scenario, the classifier may only learn how to classify edits that are reverted using the ``rollback'' method. In this case, the classifier is substantially less useful in practice, but it would show high scores in the evaluation phase for effectively ignoring all mislabeled items in the corpus.  Thus, training and testing a classifier solely based on ``rollbacked'' edits is problematic.

Rather than rely on this corpus, we applied our own strategy for identifying edits that are likely to be vandalism. First, we randomly sampled 500,000 edits saved by humans (non-bot editors) in the year 2015 in Wikidata. Next, we labeled edits that were \textit{reverted}. Next, we applied several filters to the dataset to examine cross sections of it. First, we filtered out edits that were performed by users who attained a high status in Wikidata by receiving advanced rights (including: sysop, checkuser, flood, ipblock-exempt, oversight, property-creator, rollbacker, steward, sysop, translationadmin, wikidata-staff).  Second, we filtered out edits that originated from other wikis (known as ``client edits'') and edits that merged together two Wikidata items. Finally, we were left with a set of regular edits by non-trusted users.  Next, we reviewed random samples of reverted and non-reverted edits in a few key subsets to get a sense for which of these filters could be applied when identifying vandalism.

\begin{table}
\centering
\caption{Different types of edits in a 500,000 sample}
\begin{tabular}{l|c|c} \hline
& edits & reverted \\ \hline
trusted user edit & 461176 & 1188 (0.26\%)\\ \hline
merge edit & 8241 & 38 (0.46\%) \\ \hline
client edit & 10099 & 109 (1.08\%) \\ \hline
non-trusted regular edit & 22460 & 622 (2.77\%)
\end{tabular}
\end{table}

We can safely exclude client edits since, if they are vandalism, they are vandalism to the originating wiki. Edits by trusted users are reverted at an extremely low rate, but it's still worth reviewing them, and so is reviewing merge edits. Finally, non-trusted regular edits are reverted at a high rate of 2.77\%, which is more in line with the rates seen for all edits in English Wikipedia \cite{potthast:crowdsourcing}.  To make sure that \textit{reverts} catch most of the vandalism, we manually reviewed both the reverted and non-reverted regular edits by non-trusted users.

\begin{table*}
\centering
\caption{Edits sampled for human review}
\begin{tabular}{l|c|c|c} \hline
& good & good\-faith damaging & vandalism \\ \hline
reverted merge edits & 17 & 21 & 0 \\ \hline
reverted trusted user edits & 93 & 7 & 0 \\ \hline
reverted non\-trusted regular edits & 8 & 24 & 68 \\ \hline
non\-reverted non\-trusted regular edits & 94 & 3 & 1
\end{tabular}
\end{table*}
This analysis suggests that reverted edits by non-trusted users are highly likely to be intentional vandalism (68\%) or at least damaging (92\%) and that non-reverted edits by users in this group are unlikely to be vandalism (1\%) or damaging (4\%).  Further, it appears that reverted merge edits and reverted edits by trusted users are very unlikely to be vandalism (0\% observed) – though many merges are good-faith mistakes that violate some Wikidata policy. Based on this analysis, we built a corpus of edits based on this 500,000 sample and labeled reverted regular edits by non-trusted users as \textit{True} (vandalism) and all other edits as \textit{False} (not vandalism). From this 500,000 set, we randomly split 400,000 edits for training and hyper-parameter optimization and 100,000 edits for testing. All test statistics were drawn from this 100,000 test set.

Comparing our work to that of Heindorf et al. \cite{heindorf:towards} we found that only 63\% (439) edits we identified as vandalism were reverted using the ``rollback'' method, 15\% (104) were reverted using ``restore'' and 22\% (155) were reverted using other methods.

\subsection{Feature engineering}
Before starting to build the damage detection classifier, we launched a community consultation asking Wikidata users to provide examples of common patterns of vandalism. We received around thirty patterns and examples. Community feedback was helpful to build the initial model which was launched on October 29, 2015. A second community consultation was launched for reporting possible mistakes of the initial model and more than 20 cases of false positives and false negatives were reported, which helped us improve damage detection mostly by adding proper features. In order to obtain accurately labeled data, we launched a campaign asking community members to manually label 4,283 edits. At the time of this writing, this campaign is half-way through and its data is used in this research to examine the accuracy of models and automated labels of edits that are being used in training models. Also the classifier is accessible to everyone\footnote{https://github.com/wiki-ai/wb-vandalism} which allows users and experts to comment on the algorithms and methods used.

\section{Features}
\subsection{General metrics}
\begin{itemize}
\item Number of added/removed/changed/current site links
\item Number of added/removed/changed/current labels
\item Number of added/removed/changed/current descriptions
\item Number of added/removed/changed/current statements
\item Number of added/removed/current aliases
\item Number of added/removed/current badges\footnote{Badges determine when articles linked to an item are of a high quality ("featured articles").}
\item Number of added/removed/current qualifiers
\item Number of added/removed/current references
\item Number of changed identifiers\footnote{Identifiers are properties that allow mapping an item to a corresponding item in an external knowledge bases. VIAF ids, ISBN or IMDb ids are examples of identifiers.}
\end{itemize}
\subsection{Typical vandalism patterns}
\begin{description}
\item[Proportion of Q-ids added] The proportional change to the number of \textit{items} referenced.
\item[If English label has changed] Changing the English label.
\item[Proportion of language names added]  Adding language names such as ``English'' as the values of a statement.
\item[Proportion of external links added]  Spamming Wikidata items by adding external links
\item[Is gender changed] Changing value for the gender property
\item[Is country of citizenship changed]  Changing value for country of citizenship
\item[Is member of sports team changed] Modifying statements about teams a sportsperson has played with
\item[Is date of birth changed]  Changing a person's date of birth
\item[Is image changed]  Changing the item image to an unrelated image
\item[Is image of signature changed]  Changing the signature image for people
\item[Is category of this item at Commons changed]  Changing the Wikimedia Commons category associated with an item
\item[Is official website has changed]  Changing the official website property of an item
\item[Is this item about a human]  The item is an instance of human
\item[Is this item is about a living human]  The item is an instance of human and living
\end{description}
\subsection{Typical non-vandalism patterns}
\begin{description}
\item[Is it a client edit] When a user moves a page in Wikipedia (a client of Wikidata) or deletes the page, an edit is made in Wikidata to keep the link between the projects in sync.
\item[Is it a merge] Merging -- an action that is not enabled for new users -- tends to change drastically an item's content.
\item[Is it revert, rollback, or restore] These edits are actions performed by users trying to undo vandalism.
\item[Is it creating a new item] The edit creates a new item
\end{description}
\subsection{Editor characteristics}
\begin{description}
\item[Is the user is a bot]  The edit is performed by a bot, a very common practice in Wikidata.
\item[Does the user have advanced rights]  The user is a member of the ``checkuser'', ``bureaucrat'', or ``oversight'' group and can perform advanced actions.
\item[Is the user an administrator] Administrators are advanced users with a significant amount of contributions and trusted by the community of editors
\item[Is the user a curator]  The user is a member of the``rollbacker'', ``abusefilter'', ``autopatrolled'', or ``reviewer'' group, privileges typically assigned to users with significant amounts of edits.
\item[Is the user anonymous]  The user is unregistered.
\item[Age of editor] The time between the user account registration time and the timestamp of the edit, in seconds scaled using log(age + 1).
\end{description}

\section{Evaluation}
We use three metrics to evaluate the performance of our prediction model:
\begin{itemize}
\item ROC-AUC which has been used historically in the vandalism detection literature \cite{adler:detecting, wang:got}
\item PR-AUC as suggested by Adler et al. in their more recent work \cite{adler:wikipedia}
\item Filter-rate at high recall which measures the proportion of edits that must be reviewed by Wikidata patrollers in order to for a high percentage of all vandalism to be caught.
\end{itemize}

Our inclusion of ``filter rates'' in the evaluation of the performance of vandalism classifiers is intentional since our goal is to design classifiers that can be effectively used by Wikidata patrollers.  As we improve fitness of the model, this filter-rate should increase and therefore the expected workload for patrollers should decrease: This metric directly measures theoretical changes in patroller workload.
\section{Results}
In this section, we discuss the fitness of our model against the corpus and the real-time performance of the model as exposed via ORES, our live classification service for Wikidata patrollers.
\subsection{Model fitness}
All models were tested on the exact same set of 99,222 revisions withheld during hyper-parameter optimization and training.  The table show general fitness metrics for models using different combinations of features.  As Table ~\ref{tab:model_fitness} suggests, we were only able to train marginally useful prediction models when excluding user features.  These models attained low PR-AUC values and there was no threshold that could be set on the \textit{True} probability that would allow for 75\% of \textit{reverted} edits to be identified to the exclusion of others -- resulting in a \textit{zero} filter rate.

\begin{table*}
\centering
\caption{Model fitness for different subsets of features}
\label{tab:model_fitness}
\begin{tabular}{l|c|c|c} \hline
features & ROC-AUC & PR-AUC & filter-rate \\ \hline
general & 0.777 & 0.01 & 0.936 at 0.62 recall \\  \hline
general, context & 0.803 & 0.013 & 0.937 at 0.67 recall \\  \hline
general, type, context & 0.813 & 0.014 & 0.940 at 0.68 recall \\  \hline
general, user & 0.927 & 0.387 & 0.985 at 0.86 recall \\  \hline
all & 0.941 & 0.403 & 0.982 at 0.89 recall \\  \hline
\end{tabular}
\end{table*}

\begin{figure}
\centering
\epsfig{file=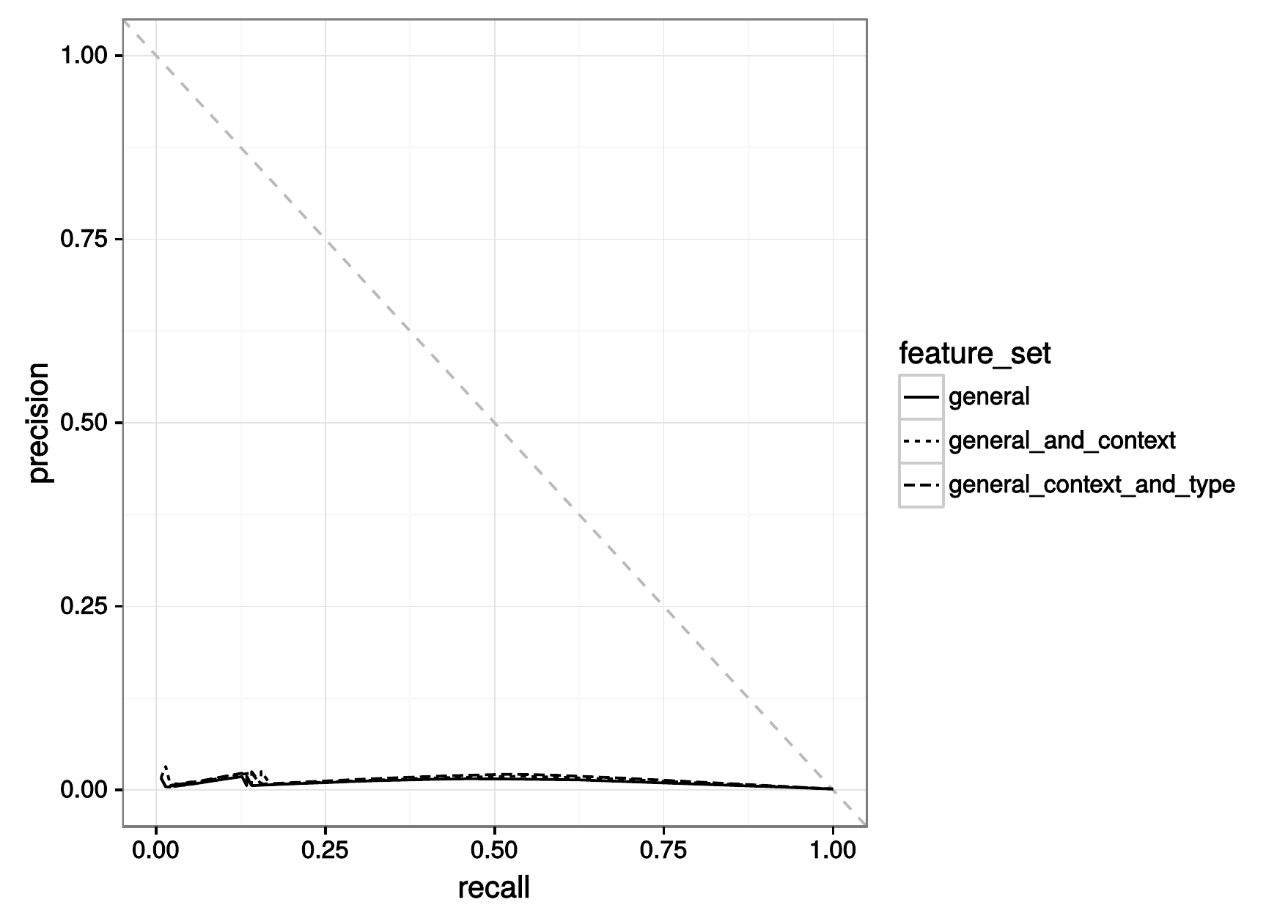, height=2in, width=3in}
\caption{\textbf{Precision/recall (no user).} The precision/recall curve for models lacking user features is plotted.}
\label{fig:precision_recall_no_user}
\end{figure}

\begin{figure}
\centering
\epsfig{file=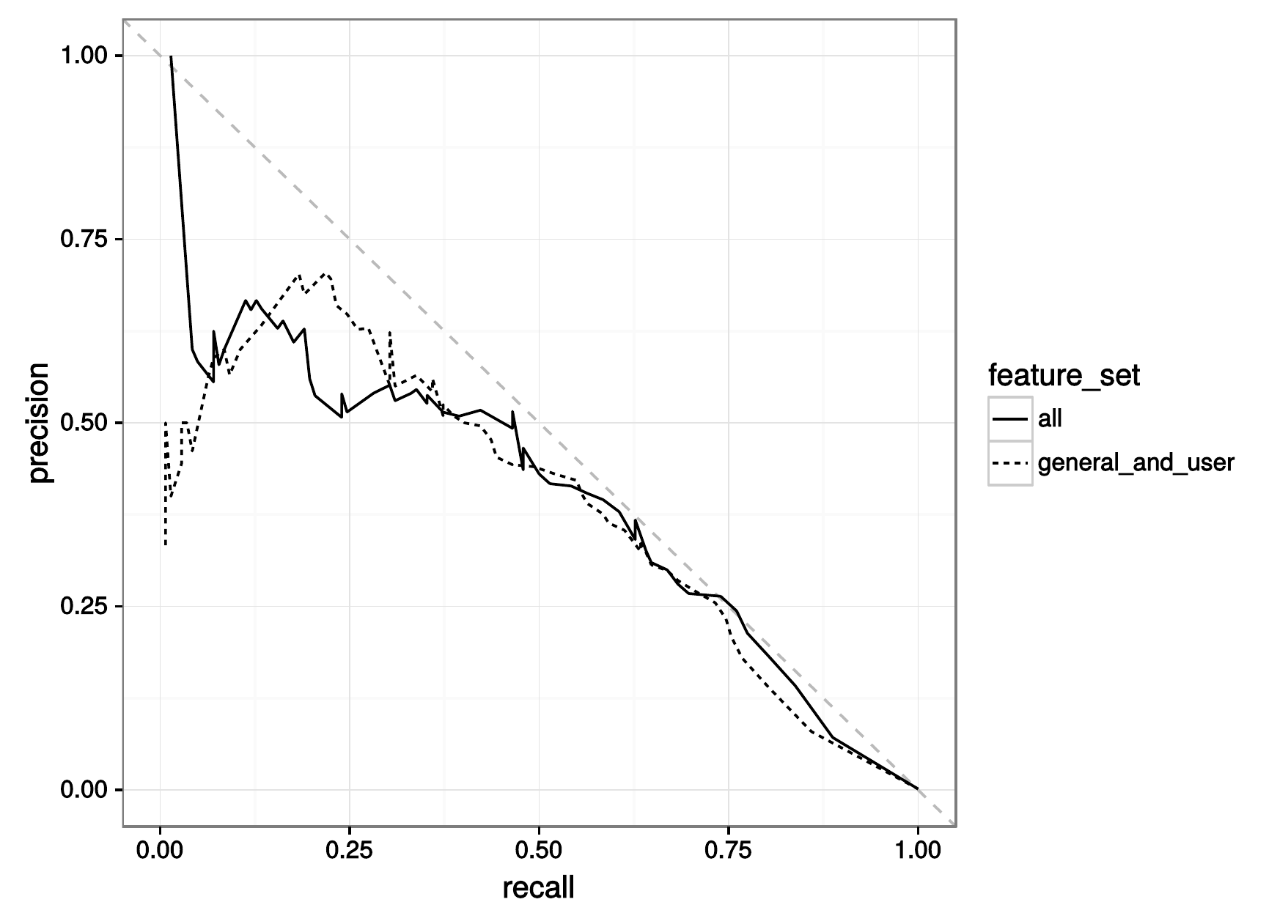, height=2in, width=3in}
\caption{\textbf{Precision/recall (user)} The precision/recall curve for models including user features is plotted.}
\label{fig:precision_recall_user}
\end{figure}

Figures ~\ref{fig:precision_recall_no_user} and ~\ref{fig:precision_recall_user} plot precision-recall curves for the two sub-feature-sets. Figure ~\ref{fig:precision_recall_no_user} visually confirms the very poor results of the classifier when no user features are included.  At the scale of the graph it is difficult to confirm any meaningfully greater than zero precision anywhere on the recall spectrum. ~\ref{fig:precision_recall_user} shows a clear difference.  For the most part, the \textit{all} features and \textit{general and user} features classifiers seem to perform comparably well across the spectrum of recall.  This suggests that the inclusion of \textit{context} and \textit{edit type} features on top of \textit{general} and \textit{user}-based features results in minor (if any) improvements.

\begin{figure}
\centering
\epsfig{file=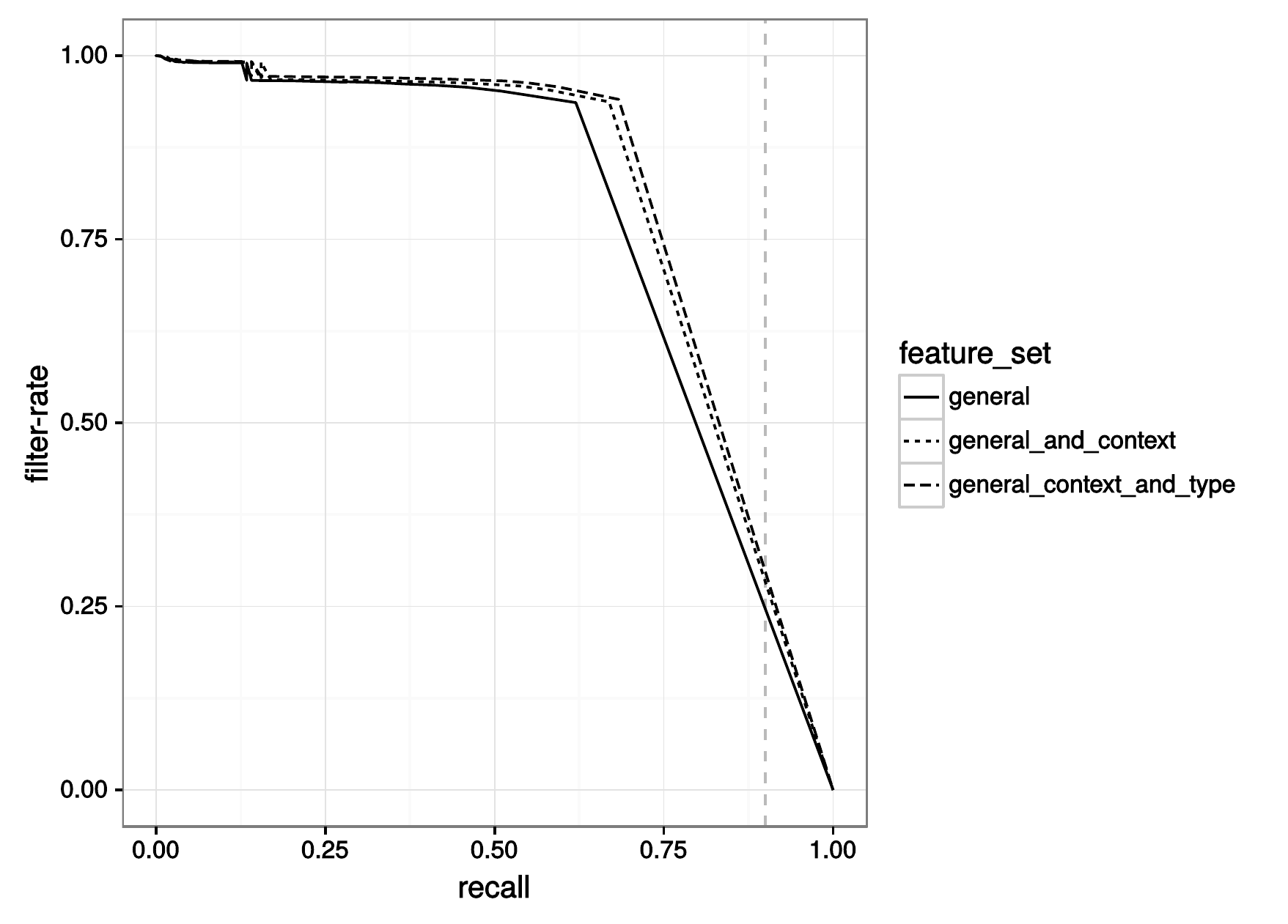, height=2in, width=3in}
\caption{\textbf{Filter-rate/recall (no user)} The filter-rate/recall curve for models lacking user features is plotted.}
\label{fig:filter_rate_no_user}
\end{figure}

\begin{figure}
\centering
\epsfig{file=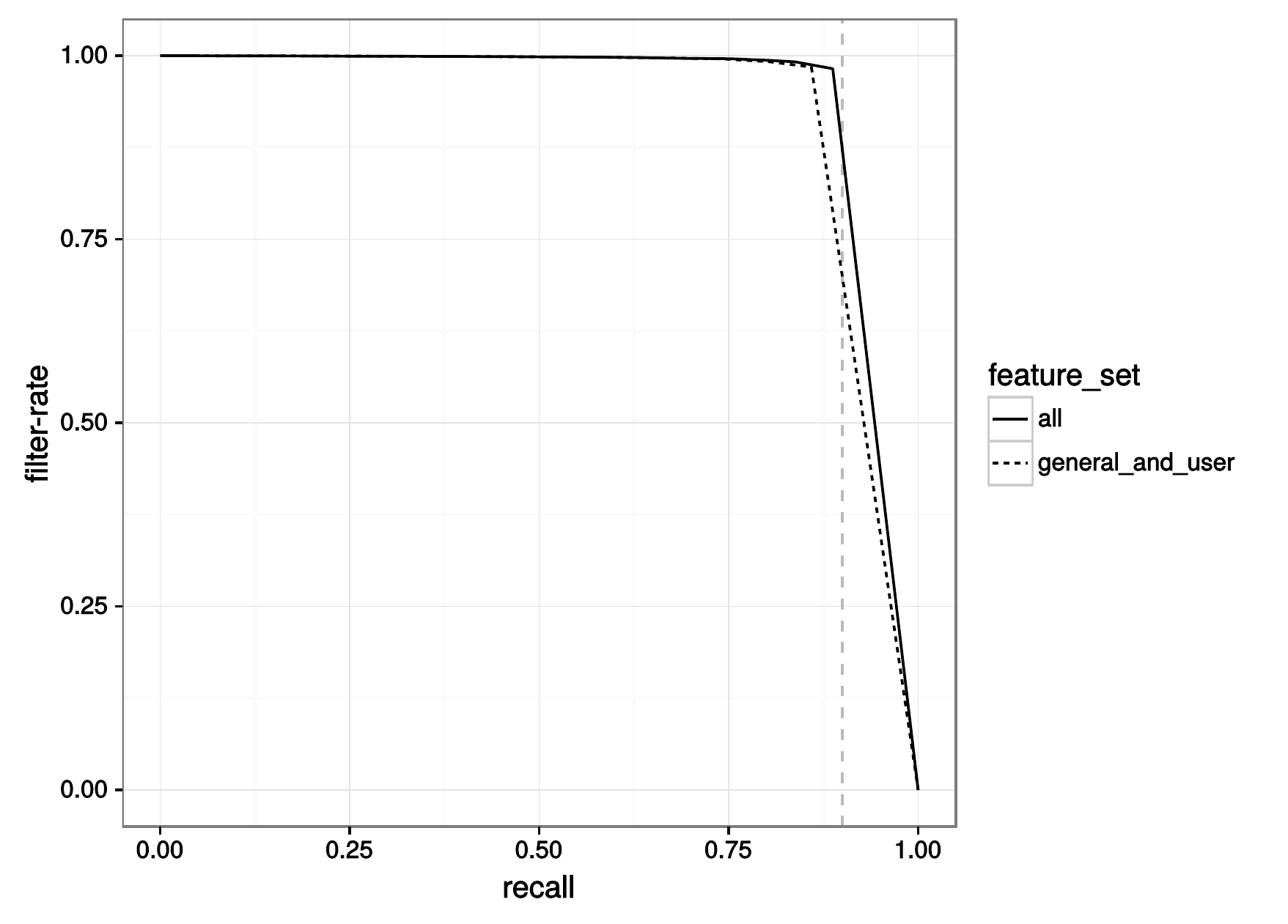, height=2in, width=3in}
\caption{\textbf{Filter-rate/recall (user)} The filter-rate/recall curve for models including user features is plotted.}
\label{fig:filter_rate_user}
\end{figure}

Figures ~\ref{fig:filter_rate_no_user} and ~\ref{fig:filter_rate_user} show the \textit{filter-rate} (which quantifies the amount of effort saved for Wikidata patrollers.  See Methods.) of the classifiers across the spectrum of recall.  Here, we can see that models that don't include \textit{user} features struggle to attain even moderate recall at any filter-rate while models that include \textit{user} features are able to attain very high filter rates up to 89\% recall.  This suggests a theoretical reduction in patrolling workload down to 1.8\% of incoming human edits, assuming that it's tolerable to let 11\% of potential vandalism to be detected by other means.

\subsection{Real-time prediction speed}
\begin{figure}
\centering
\epsfig{file=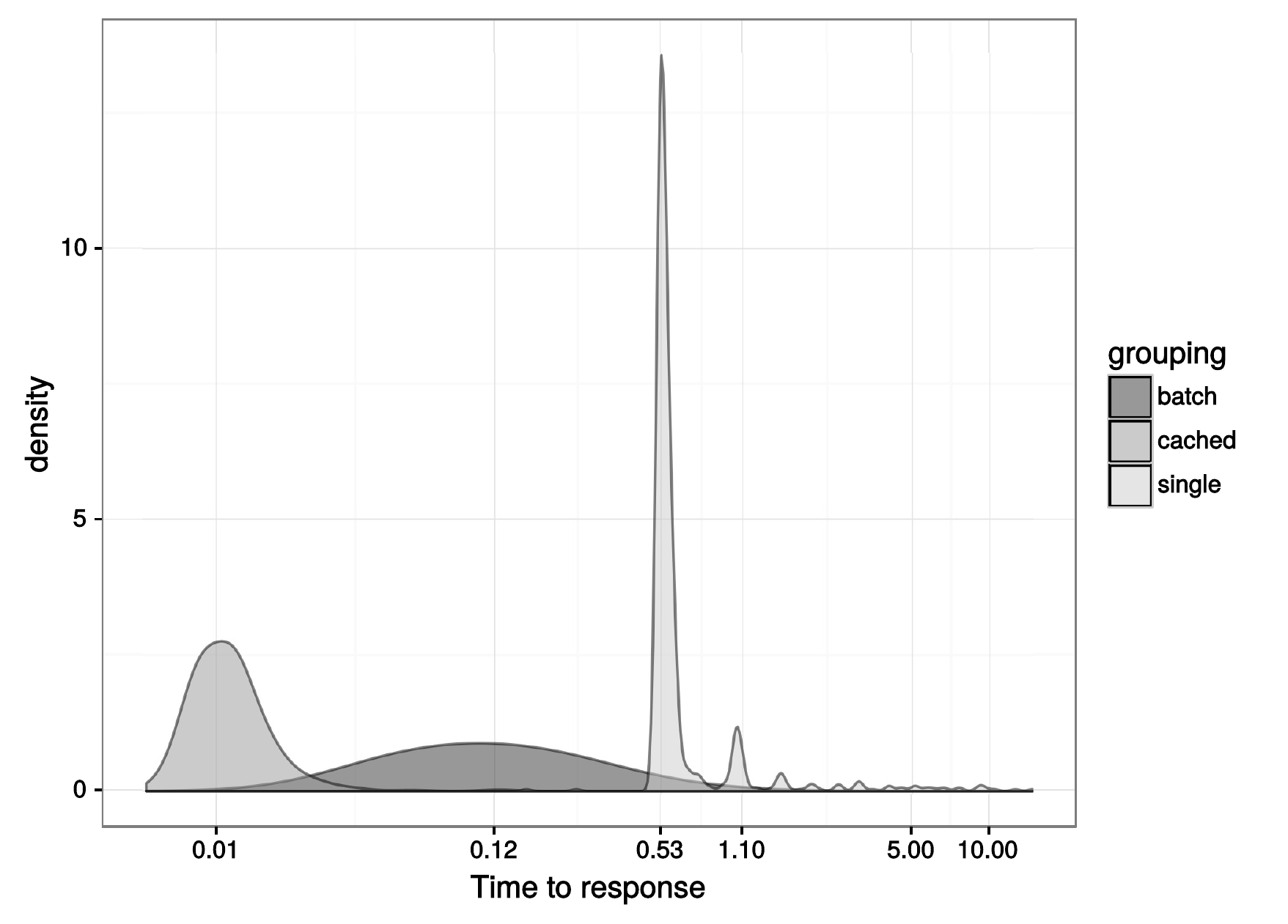, height=2in, width=3in}
\caption{Density of response timing per revision score requested from the ORES wikidata-reverted model.}
\label{fig:ores_response}
\end{figure}
Requests for the scoring of a single revision typically respond in $\sim$0.5 seconds with rare cases taking up to 2-10 seconds.  If the score has already been generated and cached, the system will generally respond in 0.01 seconds.  This is a common use-case, since we run a pre-caching service that caches scores for edits as soon as they are saved.  ORES also provides the ability to request scores in batch mode, which allows the system to gather basic data for feature extraction in batches as well.  When requesting predictions in 50 revision batches, we found that the service responds in about 0.12 seconds per revision in the batch.  As figure ~\ref{fig:ores_response} suggests, this timing varies quite widely which is likely due to rare individual edits that take a long time to score and hold back the whole batch from finishing.

\section{Limitations}

A major limitation of our model building and analysis exercise is our approach in constructing our corpus. In our analysis of which edits and reverts are likely to represent vandalism in Wikidata, we used characteristics of the edit (e.g. is it a client edit? and is it a merge edit?) and the editor (e.g. is the user in a trusted group?) to identify vandalism. These characteristics of an edit are also included as features in our prediction model. If we applied these filters in our vandalism corpus inappropriately, we would have simply trained our classifier to match these arbitrary rules we have put in place. We are confident that this is not the case generally thanks to reports from Wikidata users who have been using our live classification service. User reports generally suggest that the classifier is more effective at flagging edits that are vandalism. We also ran a follow-up qualitative analysis to help check whether our estimates of the filter-rate afforded by the ``all features'' model worked out in practice.

We randomly sampled 10,000 human edits and generated the corresponding vandalism prediction scores.  We then manually labeled (1) the highest scored edits in the dataset (100 edits at more than 93\% prediction), (2) all reverted edits in the dataset, and (3) a random samples of 100 edits for each 10\% strata of prediction weight (e.g. 30-40\%, 40-50\%, etc.)  We found only 17 (0.17\%) vandalism edits in the 10,000 set and all these vandalism observations scored 93\% or more. Only 100 of the 10,000 edits were scored 93\% or above and by reviewing this 1\% fraction of edits it was possible to catch all damaging edits. So, in this sample set, we were able to attain a 99\% filter-rate with 100\% recall by setting the threshold at 93\%.  This result looks substantially better on paper than evaluation against the test set and we think that is due to the inclusion of careful human annotation. It seems likely that more of the good edits that were mistakenly labeled as vandalism in our corpus also show up as false negatives in our test set, pushing down our apparent filter-rate and recall. While this analysis is not as robust and easy to replicate as the formal analysis we described above, we feel that it helps show that our classifier may be more useful than it appears.

These concerns and limitations call for a PAN-like dataset for Wikidata that actually uses human judgement to identify vandalism edits rather than heuristics. Lacking such a dataset, the true filter-rates and the consequent reduction in workload for patrollers can only be discovered in the context of actual work performed by patrollers. The extremely low prevalence of vandalism edits in Wikidata means that we would need extremely large numbers of labeled observations to attain a representative set of vandalism edits -- probably in the order of 100,000-1,000,000. Furthermore, and unlike vandalism in Wikipedia, labeling vandalism in Wikidata requires reviewers who are both familiar with the structure of statements in Wikidata and are able to evaluate contributions across many languages.

\section{Conclusion}
In this paper we described a straightforward method for classifying Wikidata edits as vandalism in real time by using a machine learned classification model. We show that, using this model and our prediction service, it is possible to reduce human labor involved in patrolling edits to Wikidata by nearly two orders of magnitude (98\%). At the time of writing, several tools have adopted our service and are using the prediction model to patrol incoming edits. Our analysis and a substantial part of our feature set are informed by the real-world experience of patrollers who are using this classifier to do their work.

Future work should focus on two key areas: (1) the development of a high-quality vandalism test dataset for Wikidata and (2) the development of new features for Wikidata that draw from sources of signal other than a user's status as ``untrusted''. A high-quality vandalism dataset would provide a solid basis to effectively compare the performance of prediction models without the limitations we described above and the qualitative intuitions we obtained by observing vandal fighting ``in the wild''. The development of high signal features beyond a users status are critical to design quality control processes that are fair towards users of different categories. Our classification model is strongly weighted against edits by anonymous and new contributors to Wikidata, regardless of the quality of their work.  While this may be an effective way to reduce patrollers' workload, it is likely not fair to these users that their edits be so carefully scrutinized. By increasing the fitness of this model and adding new, strong sources of signal, a classifier could help direct the patrollers attention away from good new/anonymous contributors and towards proper vandalism -- both reducing patroller workload and making Wikidata a more welcoming place for newcomers.
\section{Acknowledgments}
We would like to thank Lydia Pintscher and Abraham Taherivand from Wikimedia Deutschland. Yuvaraj Pandian from Wikimedia Foundation for operational support. Adam Wight, Helder Lima, Arthur Tilley and Gediz Aksit for their help. We also want to thanks community of Wikidata editors for providing feedback and reporting mistakes.
\bibliographystyle{abbrv}
\bibliography{refs}

\end{document}